# Predicting water flow in fully and partially saturated porous media, a new fractal-based permeability model

Nguyen Van Nghia A[1], Damien Jougnot[2], Luong Duy Thanh[1*], Phan Van Do[1],
Tran Thi Chung Thuy[1], Dang Thi Minh Hue[1], Nguyen Manh Hung[1]

[1] *Thuyloi University, 175 Tay Son, Dong Da, Hanoi, Vietnam*
[2] *Sorbonne Université, CNRS, EPHE, UMR 7619 Metis, F-75005, Paris, France*

[*]Email: thanh_lud@tlu.edu.vn

**Abstract**

Predicting the permeability of porous media in saturated and partially saturated conditions is of crucial importance in many geo-engineering areas, from water resources to vadose zone hydrology or contaminant transport predictions. Many models have been proposed in the literature to estimate the permeability from properties of the porous media such as porosity, grain size or pore size. In this study, we develop a model of the permeability for porous media saturated by one or two fluid phases with all physically-based parameters using a fractal upscaling technique. The model is related to microstructural properties of porous media such as fractal dimension for pore space, fractal dimension for tortuosity, porosity, maximum radius, ratio of minimum pore radius and maximum pore radius, water saturation and irreducible water saturation. The model is favorably compared to existing and widely used models from the literature. Then, comparison with published experimental data for both unconsolidated and consolidated samples, we show that the proposed model estimate the permeability from the medium properties very well.

## 1. Introduction

Climate change, modification of land use, groundwater and soil contamination place the society in front of vital challenges to face increasing demand for water. Understanding and predicting water flow in the critical zone, especially in aquifers and soils, is a primary need for many environmental studies and research areas (e.g., Fan et al., 2019). Permeability is one of the most crucial parameters to describe fluid flow porous media in general (e.g., Darcy, 1856; Bear, 1972). Laboratory studies have shown that the permeability depends on rock properties such as porosity, cementation, pore size, pore size distribution, pore shape and pore connectivity (e.g., Rahimi, 1977; Lis, 2019; Ghanbarian, 2020a). The permeability or relative permeability of porous media saturated by one or two phases is the key parameter that governs fluid flow in porous material and therefore plays an important role in modeling and predictions in various environmental and resources engineering. Conventionally, the permeability of porous media is determined directly in the laboratory using steady-state or unsteady-state method. Due to the complex geometric microstructure and multiscale pore structure of porous media, much effort has been devoted for predicting permeabilities. The experimental approaches for the permeability determination vary from simple measurements (e.g., Rahimi, 1977; Boulin et al., 2012; Sander et al., 2017) to indirect methods such as nuclear magnetic resonance measurements (e.g., Coates et al., 1991; Hidajat et al., 2002; Ioannidis et al., 2006), mercury injection capillary pressure measurement (Swanson, 1981), electrical conductivity measurement (e.g., Doussan and Ruy, 2009; Jougnot et al., 2010; Revil and Cathles, 1999) or spectral induced polarization measurements (e.g., Revil and Florsch, 2010; Koch et al., 2012; Revil et al., 2014; Maineult et al., 2018). In the literature, permeability prediction has been proposed through theoretical models with simplified pore geometries (e.g., Kozeny, 1927; Carman, 1938; Bear, 1972; Dullien, 1992) but also advanced schemes such as effective-medium approximations (e.g., Doyen, 1988; Richesson and Sahimi, 2019) or critical path analysis (e.g., Katz and Thompson, 1986; Hunt, 2001; Daigle, 2016; Ghanbarian, 2020a; Ghanbarian, 2020b). Additionally, explicit numerical methods such as the finite-element, finite-difference, finite-volume, lattice Boltzmann, or pore-network modeling have been applied to predict the permeability of porous materials (among many other references: Ngo and Tamma, 2001; Benzi et al., 1992; Bryant and Blunt, 1992; De Vries et al., 2017).

It is shown that porous media exhibit fractal properties and their pore spaces are statistically self-similar over several length scales (e.g., Mandelbrot, 1982; Katz and Thompson, 1985). Theory on fractal porous media has attracted much attention in different areas (e.g., Mandelbrot, 1982; Feder and Aharony, 1989). Therefore, the models based on the capillary tubes in combination with the fractal theory have been applied to study transport phenomena in both fully and partially saturated porous media (e.g., Li and Horne, 2004; Guarracino, 2007; Cai et al., 2012a,b; Liang et al., 2014; Guarracino and Jougnot, 2018; Soldi et al., 2017, 2019, 2020; Thanh et al., 2018, 2019, 2020a,b) or to study hydraulic conductivity and biological clogging in bio-amended variably saturated soils (e.g., Rosenzweig et al., 2009; Samsó et al., 2016; Carles et al., 2017). The fractal theory has already been applied to develop permeability models for porous materials. For example, Yu and Cheng (2002), Yu and Liu (2004) and Guarracino et al. (2014) developed a fractal permeability model for porous media under both saturated and partially saturated conditions. However, their models do not take into account irreducible water saturation that is very important for porous media containing small pores. Moreover, their models have not been strongly validated due to only few experimental data points used for comparison. Chen and Yao (2017) developed an improved model for the permeability estimation as an extension of Yu and Cheng (2002) and Yu and Liu (2004) by considering irreducible water saturation. Their model was verified by comparison with experimental data for natural sandstones samples whose pore size distribution is stated to be broader than that of samples such as glass/sand grains (e.g., Daigle, 2016; Ghanbarian, 2020a). Li and Horne (2004) derived a universal capillary pressure model using fractal geometry of porous media and obtained a relative permeability model using both the Purcell and the Burdine approaches, therefore obtaining model that diverges from the fractal theory. Soldi et al. (2017) and Chen et al. (2020) proposed models to describe unsaturated flow considering the hysteresis phenomena. They assumed that porous media can be represented by a bundle of capillary tubes with a periodic pattern of pore throats and pore bodies and a fractal pore size distribution. Their models have been validated using experimental data for the relative permeability and for the hysteretic saturation curves. However, Soldi et al. (2017) did not considered the variation of the capillary length with radius in their model. Chen et al. (2020) did consider that but they only focused on the relative permeability and the water retention curve rather than the intrinsic permeability. Additionally, there were not much experimental data used by Soldi et al. (2017) and Chen et al. (2020) to validate their models. Similarly, Xiao et al. (2018) obtained a model for the capillary pressure and water relative permeability in unsaturated porous rocks based on the fractal distribution of pore size and tortuosity of capillaries. It is seen that the relative permeability for water phase is a function of water saturation, porosity and fractal dimension of pore. However, Xiao et al. (2018) did not consider irreducible water saturation and did not have much experimental data to validate the model for the relative permeability. Recently, Meng et al. (2019) presented the models for both electrical conductivity and permeability based on fractal theory by introducing the critical porosity under saturated conditions. From obtained model, Meng et al. (2019) could explain the fact that the permeability of porous media could approach to zero at a nonzero percolation porosity corresponding a certain critical pore diameter. However, their model was validated by only two experimental data sets for the permeability as a function of porosity.

In this work, we develop a model for the permeability of porous media containing two fluid phases in which the fractal theory and capillary tube model are utilized. The model is related to microstructural properties of porous media such as fractal dimension for pore space, fractal dimension for tortuosity, porosity, maximum radius, ratio of minimum pore radius and maximum pore radius, water saturation and irreducible water saturation. All model parameters are physically-based parameters. The proposed model takes into account irreducible water saturation, the variation of the capillary length with radius. Then, the model for the saturated permeability $k_s$ and the relative permeability for wetting phase $k_r^w$ are validated by large published experimental data sets on 111 unconsolidated and consolidated samples. The proposed model is also compared to existing and widely used models from the literature.

## 2. Model development
### 2.1 Flow rate at the macroscale

Porous materials can be conceptualized as a bundle of tortuous capillaries following a fractal pore-size distribution (e.g., Yu and Cheng, 2002; Soldi et al., 2019; Thanh et al., 2019, 2020c). To derive analytical expressions for the permeability of porous media, we first consider a representative elementary volume (REV) of porous media as a cube with the length of $L_o$ (see Fig. 1). The pore radius $R$ of the REV varies from a minimum value $R_{min}$ to a maximum value $R_{max}$ and conforms to the fractal scaling law. Namely, the cumulative size-distribution of pores is assumed to obey the following (e.g., Yu and Cheng, 2002; Yu and Liu, 2004):

$$N(R) = \left(\frac{R_{max}}{R}\right)^{D_f} \tag{1}$$

where $N$ is the number of capillaries (whose radius $\geq R$) in the REV, $D_f$ is the fractal dimension for pore space, $0 < D_f < 2$ in two-dimensional space and $0 < D < 3$ in three dimensional space. Differentiating Eq. (1) with respect to $R$, one obtains the number of pores with radii between $R$ and $R + dR$ as

$$-dN = D_f R_{max}^{D_f} R^{-D_f - 1} dR. \tag{2}$$

The negative sign in Eq. (2) implies that the number of pores decreases when the pore radius increases.

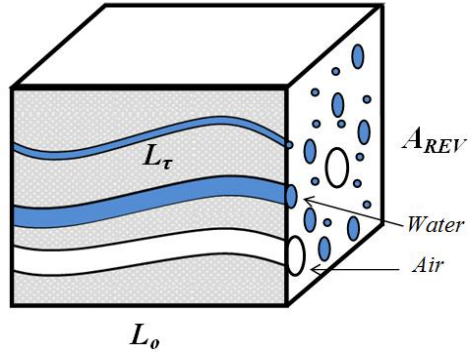

Figure 1: A porous rock model composed of a large number of parallel capillary tubes that are either saturated by water or filled by air, depending on the capillary pressure. Note that the tortuosity of the capillaries depend on their radii.

The real length of the capillary tubes $L_\tau$ along the flow direction is generally greater than the length of the porous media $L_o$ (see Fig. 1). The length $L_\tau$ is related to the pore radius $R$ as (e.g., Yu and Cheng, 2002; Yu and Liu, 2004):

$$L_\tau(R) = R^{1-D_\tau} L_o^{D_\tau}, \qquad (3)$$

where $D_\tau$ is the fractal dimension for the tortuosity ($1 \leq D_\tau \leq 2$). From Eq. (3), the tortuosity $\tau$ is determined as

$$\tau(R) = \frac{L_\tau}{L_o} = \left(\frac{L_o}{R}\right)^{D_\tau - 1}. \qquad (4)$$

The fractal dimensions $D_f$ and $D_\tau$ can be experimentally determined by a box-counting method (e.g., Yu and Cheng, 2002; Yu and Liu, 2004). In this work, they are estimated from properties of porous media. Namely, the expression for $D_f$ is given by (e.g., Yu and Cheng, 2002; Yu and Liu, 2004)

$$D_f = 2 - \frac{\ln \phi}{\ln \alpha}, \qquad (5)$$

where $\phi$ is the porosity of porous media and $\alpha = R_{min}/R_{max}$. The expression for $D_\tau$ is given by (e.g., Wei et al., 2015)

$$D_\tau = (3 - D_f) + (2 - D_f) \frac{\ln\left(\dfrac{D_f}{2F\phi^2}\right)}{\ln \phi}, \qquad (6)$$

where $F$ is the formation factor of porous media.

We consider the REV under partially saturated conditions. The REV is assumed to be initially fully saturated and then drained when it is subjected to a pressure head $h$ (m). For a capillary tube, the pore radius $R_h$ (m) is linked to a pressure head $h$ by (Jurin, 1719)

$$h = \frac{2\sigma \cos\beta}{\rho_w g R_h}, \tag{7}$$

where $\sigma$ (N/m) is the surface tension of the fluid, $\beta$ is the contact angle, $\rho_w$ (kg/m$^3$) is the fluid density and $g$ (m/s$^2$) is the acceleration due to gravity. A capillary tube of porous material becomes fully desaturated under the pressure head $h$ if its radius $R$ is larger than $R_h$ determined by Eq. (7). Hence, under the pressure head $h$, the capillaries with radii in the range between $R_{min}$ and $R_h$ will be fully saturated.

For porous media containing small pores, the irreducible water saturation can be pretty significant since water is kept in micropores (e.g., Carsel and Parrish, 1988; Jougnot et al., 2012). This amount of water is considered in this work by setting a irreducible water radius of capillaries $R_{wirr}$. Hence, we assume the following: (1) for $R_{min} \leq R \leq R_{wirr}$, the pores are filled by water that is immobile due to insufficient driving force, so it does not contribute to the water flow; (2) for $R_{wirr} < R \leq R_h$, the pores are filled by mobile water, so it contributes to the water flow; (3) for $R_h < R \leq R_{max}$, the pores are filled by air, so it does not contribute to the water flow (see Fig. 1). Note that film bound water, which adheres to the pore wall because of the molecular forces acting on the hydrophilic mineral surface, is neglected in the pores with radius larger than $R_{wirr}$. Under those assumptions, the irreducible water saturation is defined as

$$S_{wirr} = \frac{\int_{R_{min}}^{R_{wirr}} \pi R^2 L_\tau (-dN)}{\int_{R_{min}}^{R_{max}} \pi R^2 L_\tau (-dN)}. \tag{8}$$

Combining Eq. (2), Eq. (3) and Eq. (8) yields the following

$$S_{wirr} = \frac{R_{wirr}^{3-D_\tau-D_f} - R_{min}^{3-D_\tau-D_f}}{R_{max}^{3-D_\tau-D_f} - R_{min}^{3-D_\tau-D_f}}. \tag{9}$$

Similarly, water saturation is defined as

$$S_w = \frac{\int_{R_{min}}^{R_h} \pi R^2 L_\tau (-dN)}{\int_{R_{min}}^{R_{max}} \pi R^2 L_\tau (-dN)} = \frac{R_h^{3-D_\tau-D_f} - R_{min}^{3-D_\tau-D_f}}{R_{max}^{3-D_\tau-D_f} - R_{min}^{3-D_\tau-D_f}}. \tag{10}$$

Additionally, the volume flow rate in a single pore of radius $R$ (m) and length $L_\tau$ (m) is given by Poiseuille's law

$$q(R) = \frac{\rho g \pi R^4}{8\eta} \frac{\Delta h}{L_\tau}, \tag{11}$$

where $\rho$ (kg/m³), $\eta$ (Pa.s) are the density and viscosity of fluid, respectively and $\Delta h$ (m) is the pressure head drop across the REV.

The volumetric flow rate through the REV are the sum of the volumetric flow rates over all individual capillary tubes filled with water (wetting phase) and given by

$$Q_{REV}^w = \int_{R_{wirr}}^{R_h} q(R)(-dN) = \int_{R_{wirr}}^{R_h} \frac{\rho_w g \pi R^4}{8\eta_w} \frac{\Delta h_w}{L_\tau(R)} (-dN). \tag{12}$$

From Eq. (2), Eq. (3), Eq. (11) and Eq. (12), we obtain the following

$$Q_{REV}^w = \frac{\rho_w g \Delta h_w}{8\eta_w L_o^{D_\tau}} \frac{\pi D_f R_{max}^{D_f}}{3+D_\tau - D_f} (R_h^{3+D_\tau - D_f} - R_{wirr}^{3+D_\tau - D_f}). \tag{13}$$

From Eq. (8) and Eq. (10) one has

$$R_{wirr} = R_{max}[\alpha^{3-D_\tau - D_f} + S_{wirr}(1-\alpha^{3-D_\tau - D_f})]^{\frac{1}{3-D_\tau - D_f}}, \tag{14}$$

and

$$R_h = R_{max}[\alpha^{3-D_\tau - D_f} + S_w(1-\alpha^{3-D_\tau - D_f})]^{\frac{1}{3-D_\tau - D_f}}, \tag{15}$$

recall that $\alpha = R_{min}/R_{max}$.

From Eq. (13), Eq. (14) and Eq. (15), one obtains

$$\begin{aligned} Q_{REV}^w = & \frac{\pi \rho_w g \Delta h_w D_f R_{max}^{3+D_\tau}}{8\eta_w L_o^{D_\tau}} \\ & \times \left\{ [\alpha^{3-D_\tau - D_f} + S_w(1-\alpha^{3-D_\tau - D_f})]^{\frac{3+D_\tau - D_f}{3-D_\tau - D_f}} - [\alpha^{3-D_\tau - D_f} + S_{wirr}(1-\alpha^{3-D_\tau - D_f})]^{\frac{3+D_\tau - D_f}{3-D_\tau - D_f}} \right\} \end{aligned} \tag{16}$$

## 2.2 Permeability

The total volumetric flow rate through the REV can be expressed as (Buckingham, 1907)

$$Q_{REV}^w = \frac{\rho_w g}{\eta_w} k_s k_r^w \frac{\Delta h_w}{L_o} A_{REV}, \tag{17}$$

where $k_s$ (m²) is the saturated permeability, $k_r^w$ (no units) is the relative permeability for wetting phase and $A_{REV}$ is the cross sectional area of the REV.

Additionally, the porosity of the REV is calculated by

$$\phi = \frac{V_{pore}}{V_{REV}} = \frac{\int_{R_{min}}^{R_{max}} \pi R^2 L_\tau (-dN)}{L_o A_{REV}} = \frac{\pi L_o^{D_\tau - 1} D_f R_{max}^{3-D_\tau}(1-\alpha^{3-D_\tau-D_f})}{(3-D_\tau-D_f)A_{REV}}. \qquad (18)$$

Combining Eq. (16), Eq. (17) and Eq. (18), the following is obtained

$$k_s k_r^w = \frac{R_{max}^{2D_\tau}\phi}{8L_o^{2D_\tau-2}} \frac{3-D_\tau-D_f}{3+D_\tau-D_f} \frac{1}{1-\alpha^{3-D_\tau-D_f}}$$

$$\times \left\{ [\alpha^{3-D_\tau-D_f} + S_w(1-\alpha^{3-D_\tau-D_f})]^{\frac{3+D_\tau-D_f}{3-D_\tau-D_f}} - [\alpha^{3-D_\tau-D_f} + S_{wirr}(1-\alpha^{3-D_\tau-D_f})]^{\frac{3+D_\tau-D_f}{3-D_\tau-D_f}} \right\} \qquad (19)$$

Using Eq. (19) and invoking $k_r^w = 1$ at $S_w = 1$, we obtain the following

$$k_s = \frac{R_{max}^{2D_\tau}\phi}{8\eta L_o^{2D_\tau-2}} \frac{3-D_\tau-D_f}{3+D_\tau-D_f} \frac{\left\{1 - [\alpha^{3-D_\tau-D_f} + S_{wirr}(1-\alpha^{3-D_\tau-D_f})]^{\frac{3+D_\tau-D_f}{3-D_\tau-D_f}}\right\}}{1-\alpha^{3-D_\tau-D_f}}, \qquad (20)$$

and

$$k_r^w = \frac{\left\{[\alpha^{3-D_\tau-D_f} + S_w(1-\alpha^{3-D_\tau-D_f})]^{\frac{3+D_\tau-D_f}{3-D_\tau-D_f}} - [\alpha^{3-D_\tau-D_f} + S_{wirr}(1-\alpha^{3-D_\tau-D_f})]^{\frac{3+D_\tau-D_f}{3-D_\tau-D_f}}\right\}}{\left\{1 - [\alpha^{3-D_\tau-D_f} + S_{wirr}(1-\alpha^{3-D_\tau-D_f})]^{\frac{3+D_\tau-D_f}{3-D_\tau-D_f}}\right\}}. \qquad (21)$$

Eq. (20) can be written as

$$k_s = \frac{R_{max}^2 \phi(3-D_\tau-D_f)}{8(\tau^{eff})^2(3+D_\tau-D_f)} \frac{\left\{1 - [\alpha^{3-D_\tau-D_f} + S_{wirr}(1-\alpha^{3-D_\tau-D_f})]^{\frac{3+D_\tau-D_f}{3-D_\tau-D_f}}\right\}}{1-\alpha^{3-D_\tau-D_f}}, \qquad (22)$$

where $\tau^{eff}$ is given by

$$\tau^{eff} = \left(\frac{L_o}{R_{max}}\right)^{D_\tau-1} \qquad (23)$$

and that is defined as the effective tortuosity of the porous medium as inferred from Eq.(4) (Thanh et al., 2019, 2020c).

The length of the cubic REV is related to the cross-section area of the REV by

$$L_o^2 = A_{REV}, \qquad (24)$$

From Eq. (18), Eq. (23) and Eq. (24), one has

$$\tau^{eff} = \left[ \frac{1-\alpha^{3-D_\tau-D_f}}{\phi} \frac{\pi D_f}{3-D_\tau-D_f} \right]^{\frac{D_\tau-1}{3-D_\tau}}. \qquad (25)$$

Eq. (21) and Eq. (22) are the key contributions of this work. These equations show that the saturated permeability and the relative permeability for wetting phase are functions of microstructural parameters of porous media ($D_f$, $D_\tau$, $\phi$, $\alpha$, $R_{max}$), $S_w$ and $S_{wirr}$. Therefore, the proposed model provides an insight into the dependence of the saturated permeability ($k_s$) and the relative permeability ($k_r^w$) on the microstructural parameters of the porous media and it may reveal more mechanisms affecting the $k_s$ and $k_r^w$ than other models. In particular, the proposed model contains physically-based parameters and that is different from some other models in literature (see Table 2) with empirical parameters such as *m* in the RC model, *b* that is normally taken as 180 in the KC model, *a* and *m* in the RGPZ model or *c* that is normally taken as 72.2 in the CPA model.

In case of $R_{max} \gg R_{min}$ ($\alpha \to 0$), that is normally acceptable for porous rocks (see Guarracino, 2007; Soldi et al., 2019) and the negligible irreducible water saturation $S_{wirr} = 0$, Eq. (21) and Eq. (22), respectively, becomes

$$k_r^w = S_w^{\frac{3+D_\tau-D_f}{3-D_\tau-D_f}} \qquad (26)$$

and

$$k_s = \frac{R_{max}^2 \phi (3-D_\tau-D_f)}{8(\tau^{eff})^2 (3+D_\tau-D_f)} \qquad (27)$$

It is seen that Eq. (26) is similar to the power law of the Burdine-Brooks-Corey model (Brooks & Corey, 1964; Burdine, 1953) that is given by (Ghanbarian et al., 2017a):

$$k_r^w = S_w^{\mu+1+\frac{2}{\lambda}} \qquad (28)$$

where $\mu$ is the empirical tortuosity-connectivity exponent and $\lambda$ is the pore size distribution index. Obviously, the number of parameters in Eq. (26) ($S_w$, $D_f$ and $D_\tau$) is the same as that in Eq. (28) ($S_w$, $\mu$ and $\lambda$). Eq. (27) has five parameters ($R_{max}$, $\tau^{eff}$, $\phi$, $D_f$ and $D_\tau$) that are comparable with the number of parameters in some other models in literature as reported in Table 2 (e.g., three parameters of *d*, *m* and *F* in the RC model, four parameters of *d*, $\phi$, *a* and *m* in the RGPZ model, three parameters of $d_c$, *c* and *F* in the CPA model).

If one does not consider the variation of tortuosity with the capillary radius then $D_\tau = 1$ and Eq. (26) becomes

$$k_r^w = S_w^{\frac{4-D_f}{2-D_f}}. \qquad (29)$$

Eq. (29) is the same as that in Soldi et al. (2019).

## 3. Results and discussion

To predict $k_r^w$ from Eq. (21) and $k_s$ from Eq. (22), one needs to know parameters: $D_f$, $D_\tau$, $R_{max}$, $\phi$, $F$, $\alpha$, $S_w$ and $S_{wirr}$. Note that these model parameters are physically-based parameters. For example, the fractal dimension of the capillary size distribution $D_f$ represents the heterogeneity of the porous medium. The greater the fractal dimension, the more heterogeneous the porous media (e.g.., Li and Horne, 2004 ; Othman et al., 2010 ; Zainaldin et al., 2017). The fractal dimension of the toruosity $D_\tau$ represents the extent of convolutedness of capillary pathways for fluid flow through porous media; $D_\tau = 1$ corresponds to straight capillary paths and a higher value of $D_\tau$ corresponds to a highly tortuous capillary in porous media (e.g., Feng and Yu, 2007 ; Cai and Yu, 2011). Values of $D_f$ and $D_\tau$ can be determined by the box-counting method (e.g., Yu and Cheng, 2002; Yu and Liu, 2004; Othman et al., 2010). The other parameters such as $R_{max}$ $\phi$, $F$, $\alpha$, $S_w$ and $S_{wirr}$ can be determined in the lab. For example, the porosity $\phi$ can be measured by different methods such as the mercury porosimetry, helium pycnometry, image analysis and water absorption, among other ones (e.g., Andreola et al., 2000; Nnaemeka, 2010). The grain diameter $d$ can be determined by techniques such as the sieve analysis, the laser diffraction, the microscopy technique and others (e.g., Li et al., 2005; Abbireddy and Clayton, 2009). The formation factor $F$ can be measured by an approach presented by Jouniaux et al., 2000 or Vinogradov et al., 2010, for example. In the context of a bundle of capillary tubes model, $R_{min}$ and $R_{max}$ correspond to the sizes of pores invaded by the nonwetting phase at the maximum and minimum values of capillary pressure. Therefore, they can be estimated by measuring the maximum capillary pressure and the minimum capillary pressure, respectively, then using the Young–Laplace equation (e.g., Ghanbarian et al., 2017b). Additionally, Daigle, 2016 determined $R_{min}$, $R_{max}$ and therefore $\alpha$ from the micro-CT images and the nuclear magnetic resonance (NMR) measurements. He combined the micro-CT and the NMR data to provide a continuous pore size distribution in pores and therefore obtained $R_{min}$, $R_{max}$ and $\alpha$. Note that $S_{wirr}$ can be obtained from the soil water retention curves that are measured by methods such as the pressure plate, tensiometers, or pressure membranes, for example (e.g., Lourenço et al., 2007; Abeykoon et al., 2017).

If the pore size distribution is unknown, the $R_{max}$ for non consolidated granular media can be estimated by following (Cai et al., 2012a; Liang et al., 2014):

$$R_{max} = \frac{d}{8}\left[\sqrt{\frac{2\phi}{1-\phi}} + \sqrt{\frac{\phi}{1-\phi}} + \sqrt{\frac{\pi}{4(1-\phi)}} - 1\right], \quad (30)$$

Based on the published work from Ghanbarian (2020a) or Revil and Cathles (1999), we use $S_{wirr} = 0$ for the intrinsic permeability to simplify parameter optimization. For the relative permeability estimation, $S_{wirr}$ is obtained through an optimization procedure. Namely, the optimization of parameters is based on data fitting and then calculating the root-mean-square error (RMSE). Model parameters are then determined by seeking a minimum RMSE through the "fminsearch" function in the MATLAB. In this work, we use the "fminsearch" function to optimize parameters of $\alpha$, $D_f$ and $D_\tau$ for the intrinsic permeability and to optimize parameters of $\alpha$, $D_f$, $D_\tau$ and $S_{wirr}$ for the relative permeability.

### 3.1 Saturated permeability

To study the model sensitivity with model parameters such as $\phi$, $\alpha$, $S_{wirr}$ and $D_\tau$, from Eq. (22) we predict the variation of $k_s$ with porosity as shown in Fig. 2 (a) and with irreducible water saturation as shown in Fig. 2 (b). Fig. 2 (a) is obtained with $S_{wirr} = 0$, $D_\tau = 1.1$, $R_{max} = 40$ μm and three values of $\alpha$ (0.0001, 0.001 and 0.01). Fig. 2 (b) is obtained with $\alpha = 0.01$, $\phi = 0.2$, $R_{max} = 40$ μm and three values of $D_\tau$ (1.1, 1.15 and 1.2). Note that the input parameters for Fig. 2 are normally in the range reported in literature for porous media. For example, $\alpha$ is commonly between 0.0001 and 0.01 (e.g., Wei et al., 2015; Thanh et al., 2020c); $D_\tau$ is normally reported to be around 1.1 (e.g., Chen et al., 2020) and $R_{max}$ is reported to be tens of micrometer in geological media (e.g., Hu et al., 2017). It is seen that that the permeability $k_s$ is sensitive to $\phi$, $\alpha$, $S_{wirr}$ and $D_\tau$. Namely, $k_s$ increases with increasing porosity as expected by other models (e.g., Kozeny, 1927; Revil and Cathles, 1999) and with increasing $\alpha$ as predicted by Xu and Yu (2008). Additionally, $k_s$ decreases with an increase of $S_{wirr}$. This is attributed to the fact that the larger $S_{wirr}$ causes the total flow rate of wetting phase smaller and therefore the permeability decreases. It is also seen that $k_s$ decreases with an increase of $D_\tau$. The reason is that when $D_\tau$ increases, the flow pathways are more tortuous, causing more resistance for flow and lower the permeability of porous media. It should be noted that in Fig. 2, $D_f$ is estimated from Eq. (5) with the knowledge of $\phi$ and $\alpha$.

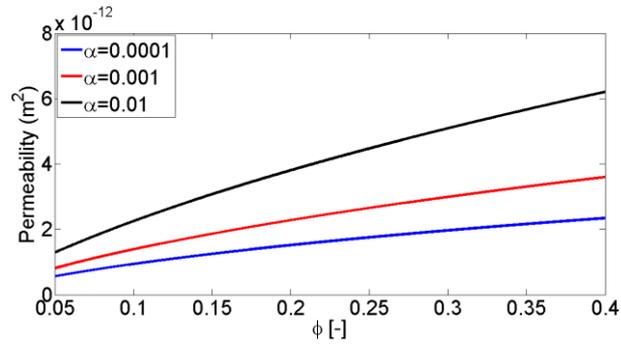

(a)

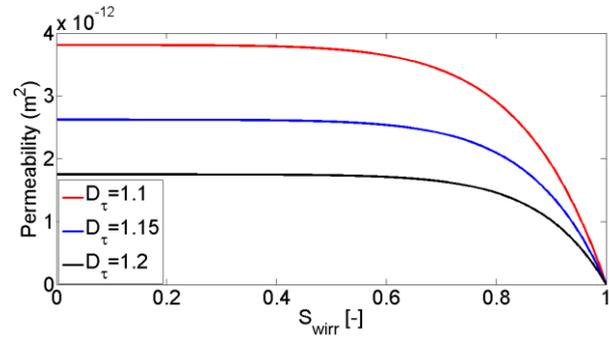

(b)

Figure 2: Sensitivity analysis of the model: (a) variation of the $k_s$ with porosity for three values of $\alpha$ (0.0001, 0.001 and 0.01) at $S_{wirr} = 0$, $D_\tau = 1.1$, $R_{max} = 40$ μm and (b) variation of the $k_s$ with irreducible water saturation $S_{wirr}$ for three values of $D_\tau$ (1.1, 1.15 and 1.2) at $\alpha = 0.01$, $\phi = 0.2$, $R_{max} = 40$ μm.

Table 1: Properties of the glass bead and sand packs used in this work. Symbols $d$, $\phi$, $F$, $\alpha$, $D_f$, $D_\tau$ and $k_s$ stand for the grain diameter, porosity, formation factor, ratio of minimum and maximum radius, fractal dimension for pore space, fractal dimension for the tortuosity and permeability of samples, respectively. Note that superscripts *, a, o and p stand for measured quantities, estimated ones from Archie (1942), optimized ones by the "fminsearch" function in Matlab and predicted ones from Eq. (5) and Eq. (6), respectively.

| Pack | $d^*$ ($\mu$m) | $\phi^*$ (no units) | $F^{*,a}$ (no units) | $k_s^*$ $10^{-12}$ (m$^2$) | $\alpha$ | $D_f$ | $D_\tau$ | Source | Shown in |
|---|---|---|---|---|---|---|---|---|---|
| Glass bead | 56 | 0.4 | 3.3* | 2.0 | 0.0103° | 1.803° | 1.05° | Bolève et al. (2007) | Fig. 3(a) |
| | 72 | 0.4 | 3.2* | 3.1 | | 1.789p from Eq. (5) | 1.11p from Eq. (6) | | |
| | 93 | 0.4 | 3.4* | 4.4 | | | | | |
| | 181 | 0.4 | 3.3* | 27 | | | | | |
| | 256 | 0.4 | 3.4* | 56 | | | | | |
| | 512 | 0.4 | 3.4* | 120 | | | | | |
| | 3000 | 0.4 | 3.6* | 14000 | | | | | |
| Glass bead | 75 | 0.43 | 3.55a | 5.3 | 0.0090° | 1.900° | 1.14° | Johnson et al. (1986b) | Fig. 3(a) |
| | 110 | 0.41 | 3.81a | 8.6 | | 1.814p from Eq (5) | 1.11p from Eq (6) | | |
| | 500 | 0.41 | 3.81a | 174.6 | | | | | |
| Glass bead | 11.5 | 0.41 | 3.81a | 0.11 | 0.0091° | 1.904° | 1.13° | Chauveteau and Zaitoun (1981) | Fig. 3(a) |
| | 15 | 0.41 | 3.81a | 0.21 | | 1.809p from Eq. (5) | 1.11p from Eq. (6) | | |
| | 25 | 0.41 | 3.81a | 0.66 | | | | | |
| | 45 | 0.41 | 3.81a | 2.4 | | | | | |
| | 90 | 0.4 | 3.95a | 8.4 | | | | | |
| | 225 | 0.4 | 3.95a | 36.0 | | | | | |
| | 450 | 0.4 | 3.95a | 137 | | | | | |
| Glass bead | 20 | 0.4009 | 3.90* | 0.24 | 0.0095° | 1.793° | 1.15° | Glover et al. (2006) | Fig. 3(a) |
| | 45 | 0.3909 | 4.02* | 1.6 | | 1.797p from Eq. (5) | 1.12p from Eq. (6) | | |
| | 106 | 0.3937 | 4.05* | 8.1 | | | | | |
| | 250 | 0.3982 | 3.98* | 50.5 | | | | | |
| | 500 | 0.3812 | 4.09* | 186.8 | | | | | |
| | 1000 | 0.3954 | 3.91* | 709.9 | | | | | |
| | 2000 | 0.3856 | 4.14* | 2277.3 | | | | | |
| | 3350 | 0.3965 | 3.93* | 7706.9 | | | | | |
| Glass bead | 3000 | 0.398 | 4.21* | 4892 | 0.0092° | 1.736° | 1.26° | Glover and Walker (2009) | Fig. 3(a) |
| | 4000 | 0.385 | 4.38* | 6706 | | 1.898p from Eq. (5) | 1.12p from Eq. (6) | | |
| | 5000 | 0.376 | 4.65* | 8584 | | | | | |
| | 6000 | 0.357 | 5.31* | 8262 | | | | | |
| | 256 | 0.399 | 4.01* | 41.2 | | | | | |
| | 512 | 0.389 | 4.36* | 164 | | | | | |
| | 181 | 0.382 | 4.39* | 18.6 | | | | | |
| Glass bead | 115 | 0.366 | 4.09* | 8.8 | 0.0097° | 1.753° | 1.11° | Kimura (2018) | Fig. 3(b) |
| | 136 | 0.364 | 4.20* | 10.7 | | 1.780p from Eq. (5) | 1.11p from Eq. (6) | | |
| | 162 | 0.363 | 4.13* | 18.3 | | | | | |
| | 193 | 0.364 | 4.04* | 26.7 | | | | | |
| | 229 | 0.362 | 4.20* | 33.0 | | | | | |
| | 273 | 0.358 | 4.17* | 51.0 | | | | | |
| | 324 | 0.358 | 4.15* | 67.4 | | | | | |
| | 386 | 0.356 | 4.36* | 102.1 | | | | | |
| | 459 | 0.358 | 4.30* | 134.3 | | | | | |
| | 545 | 0.36 | 4.06* | 246.2 | | | | | |
| | 648 | 0.358 | 4.18* | 299 | | | | | |
| | 771 | 0.357 | 4.29* | 510.4 | | | | | |
| | 917 | 0.356 | 4.15* | 611.9 | | | | | |
| Silica sand | 115 | 0.379 | 4.02* | 7.0 | 0.0066° | 1.789° | 1.15° | | |
| | 136 | 0.378 | 4.27* | 10.9 | | 1.808p from | 1.11p from | | |
| | 162 | 0.378 | 4.21* | 16.6 | | | | | |

| Material | | | | | | | Reference | |
|---|---|---|---|---|---|---|---|---|
| | 193 | 0.378 | 4.16* | 20.0 | | Eq. (5) | Eq. (6) | |
| | 229 | 0.38 | 4.24* | 27.5 | | | | |
| | 273 | 0.38 | 4.15* | 45.4 | | | | |
| | 324 | 0.38 | 4.07* | 70.5 | | | | |
| | 386 | 0.38 | 4.12* | 89.9 | | | | |
| | 459 | 0.381 | 4.17* | 133.7 | | | | |
| | 545 | 0.383 | 4.09* | 189.6 | | | | |
| | 648 | 0.385 | 4.12* | 270.8 | | | | |
| | 771 | 0.388 | 4.10* | 391.7 | | | | |
| | 917 | 0.389 | 3.95* | 558.6 | | | | |
| Fujikawa sand | 162 | 0.442 | 3.75* | 14.4 | 0.0093 o | 1.757 o | 1.20 o | |
| | 229 | 0.421 | 3.83* | 27.8 | | 1.814 p | 1.12 p | |
| | 273 | 0.419 | 3.79* | 42.9 | | from | from | |
| | 324 | 0.416 | 3.88* | 56.5 | | Eq. (5) | Eq. (6) | |
| | 386 | 0.413 | 3.90* | 81.8 | | | | |
| | 459 | 0.414 | 3.93* | 123.8 | | | | |
| | 545 | 0.415 | 3.92* | 176.8 | | | | |
| | 648 | 0.415 | 3.91* | 234.6 | | | | |
| Sand | 150 | 0.45 | 3.92* | 6.7 | 0.0092 o | 1.612 o | 1.31 o | Biella et al. (1983) | Fig. 3(c) |
| | 300 | 0.43 | 4.10* | 49.2 | | 1.801 p | 1.13 p | | |
| | 500 | 0.40 | 4.05* | 107.7 | | from | from | | |
| | 800 | 0.41 | 4.29* | 205.1 | | Eq. (5) | Eq. (6) | | |
| | 1300 | 0.40 | 4.20* | 810.2 | | | | | |
| | 1800 | 0.39 | 4.31* | 1261.4 | | | | | |
| | 2575 | 0.37 | 4.77* | 2563.8 | | | | | |
| | 3575 | 0.38 | 4.88* | 5127.6 | | | | | |
| | 4500 | 0.37 | 4.64* | 5640.4 | | | | | |
| | 5650 | 0.37 | 4.70* | 8204.2 | | | | | |
| | 7150 | 0.37 | 4.70* | 12306.3 | | | | | |
| Sand | 192 | 0.383 | 4.22 a | 21.4 | 0.0095 o | 1.780 o | 1.14 o | Moghadasi et al. (2004) | Fig. 3(c) |
| | 265 | 0.383 | 4.22 a | 60.3 | | 1.794 p | 1.12 p | | |
| | 410 | 0.384 | 4.20 a | 121 | | from | from | | |
| | 1000 | 0.385 | 4.19 a | 727 | | Eq. (5) | Eq. (6) | | |
| Quartz sand | 180 | 0.47 | 3.77* | 17.6 | 0.007 o | 1.460 o | 1.41 o | Koch et al. (2012) | Fig. 3(d) |
| | 270 | 0.45 | 3.55* | 53.1 | | 1.835 p | 1.10 p | | |
| | 660 | 0.47 | 3.25* | 129 | | from | from | | |
| | 180 | 0.48 | 3.14* | 20.8 | | Eq. (5) | Eq. (6) | | |
| | 240 | 0.49 | 3.40* | 33.0 | | | | | |
| | 320 | 0.49 | 3.26* | 67.5 | | | | | |
| | 500 | 0.49 | 3.12* | 171 | | | | | |
| | 680 | 0.48 | 3.10* | 280 | | | | | |
| | 870 | 0.49 | 3.34* | 394 | | | | | |
| | 180 | 0.39 | 4.12* | 11.1 | | | | | |
| | 270 | 0.39 | 3.75* | 24.0 | | | | | |
| | 660 | 0.41 | 3.97* | 75.0 | | | | | |
| | 180 | 0.40 | 3.23* | 11.7 | | | | | |
| | 240 | 0.40 | 3.55* | 19.8 | | | | | |
| | 320 | 0.42 | 3.64* | 38.1 | | | | | |
| | 500 | 0.42 | 3.52* | 105.0 | | | | | |
| | 680 | 0.42 | 3.36* | 196.0 | | | | | |
| | 870 | 0.41 | 3.63* | 256.0 | | | | | |
| Glass bead | 1.05 | 0.411 | 3.80 a | 0.00057 | 0.009 | 1.800 p | 1.12 p | Glover and Dery (2010) | Fig. 4 |
| | 2.11 | 0.398 | 3.98 a | 0.00345 | | from | from | | |
| | 5.01 | 0.380 | 4.27 a | 0.0181 | | Eq. (5) | Eq. (6) | | |
| | 11.2 | 0.401 | 3.94 a | 0.0361 | | | | | |
| | 21.5 | 0.383 | 4.22 a | 0.228 | | | | | |
| | 31 | 0.392 | 4.07 a | 0.895 | | | | | |
| | 47.5 | 0.403 | 3.91 a | 1.258 | | | | | |
| | 104 | 0.394 | 4.04 a | 6.028 | | | | | |
| | 181 | 0.396 | 4.01 a | 21.53 | | | | | |

| | | | | | | | | |
|---|---|---|---|---|---|---|---|---|
| 252 | 0.414 | 3.75[a] | 40.19 | | | | | |
| 494 | 0.379 | 4.29[a] | 224 | | | | | |
| 990 | 0.385 | 4.19[a] | 866.7 | | | | | |
| Average | | | | 0.0090 | | | | |

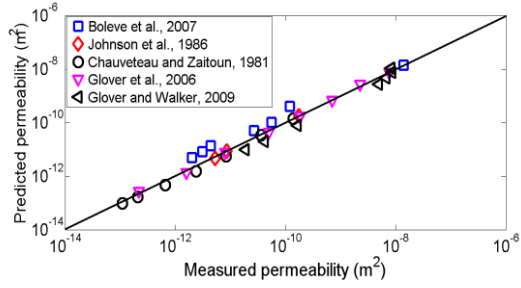

(a)

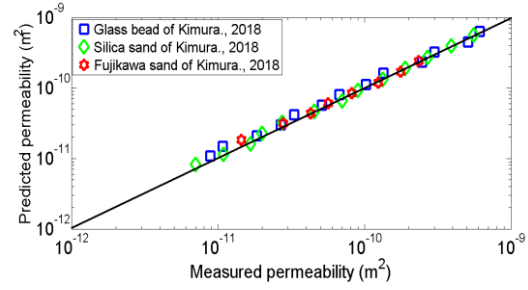

(b)

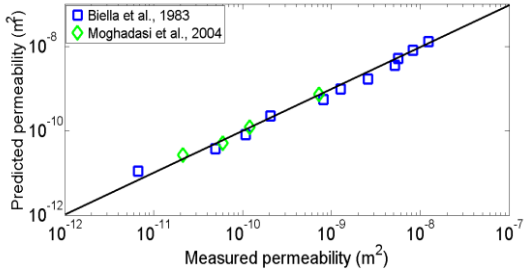

(c)

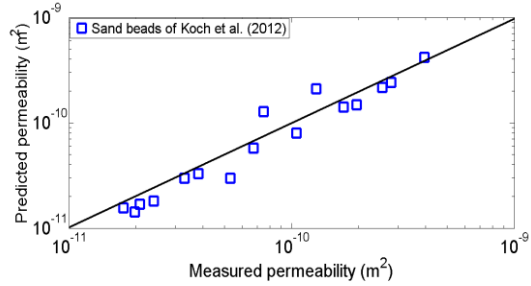

(d)

Figure 3: Comparison between measured permeability reported in literature and the one estimated by Eq. (22) in which sample properties are given in Table 1 with $S_{wirr} = 0$: (a) glass beads (data from Bolève al., 2007; Johnson et al., 1986b; Chauveteau and Zaitoun, 1981; Glover et al., 2006; Glover and Walker, 2009), (b) glass beads and silica sands (data from Kimura, 2018), (c) silica sands (data from Biella et al., 1983; Moghadasi et al., 2004), and (d) sand grains (data from Koch et al., 2012). The solid lines represent the 1:1 line.

Figure 3 shows the comparison between the predicted permeability by Eq. (22) and the measured permeability for 111 uniform packs from different sources: (a) for glass beads (data from Bolève et al., 2007; Johnson et al., 1986b; Chauveteau and Zaitoun, 1981; Glover et al., 2006; Glover and Walker, 2009), (b) for glass beads and silica sands (data from Kimura, 2018), (c) for silica sands (data from Biella et al., 1983; Moghadasi et al., 2004), and (d) for sand grains (data from Koch et al., 2012). The sample properties ($d$, $\phi$ and $F$) as well as the measured $k_s$ are summarized in Table 1. The formation factor $F$ is not available for the samples reported by Johnson et al. (1986b), Chauveteau and Zaitoun (1981), Moghadasi et al. (2004) and Glover and Dery (2010). Therefore, we estimate $F$ from $\phi$ by the relation $F = \phi^{-m}$ (Archie, 1942) with $m = 1.5$ for spherical beads (e.g.,

Sen et al., 1981). Model parameters of $\alpha$, $D_f$, $D_\tau$ are optimized using the "fminsearch" function in Matlab for Fig. 3 as mentioned above and are shown in Table 1 with the superscript [o]. $R_{max}$ is determined from Eq. (30) with the knowledge of $d$ and $\phi$ (see columns 2 and 3 in Table 1). As seen that the average optimized value for $\alpha$ is around 0.009. That value is in good agreement with $\alpha = 0.01$ that has been effectively applied for unconsolidated samples such as sand packs or glass beads (e.g., Cai et al., 2012a; Liang et al., 2015; Thanh et al., 2018, 2019). Additionally, we also predict $D_f$ from $\phi$ and the optimized value $\alpha$ using Eq. (5) and predict $D_\tau$ from $D_f$, $\phi$ and $F$ using Eq. (6). The predicted values are shown in Table 1 with the superscript [p]. It is seen that the predicted values for $D_f$, $D_\tau$ are close to the optimized ones by the "fminsearch" (average difference by 4%). The comparison in Fig. 3 shows that the model predictions are in quite good agreement with experimental data.

Table 2: Some of the models for the grain-size-based permeability estimation. Recall that $d$ is the grain diameter, $\phi$ is the porosity, $F$ is the formation factor, $m$ and $a$ are parameters taken as 1.5 and 8/3 for spherical grain samples (e.g., Sen et al., 1981; Glover et al., 2006). Note that $d_c$ and $c$ in the CPA model are the critical pore diameter and a constant coefficient equal to 72.2 (e.g., Ghanbarian, 2020a).

| Model | Equation | Reference |
|---|---|---|
| RC model | $k_s = \dfrac{d^2}{8m^2 F(F-1)^2}$ | Revil and Cathles (1999); Koch et al. (2012) |
| KC model | $k_s = \dfrac{d^2 \phi^3}{180(1-\phi)^2}$ | Kozeny (1927); Revil and Cathles (1999) |
| RGPZ model | $k_s = \dfrac{d^2 \phi^{3m}}{4am^2}$ | Glover et al. (2006) |
| XY model | $k_s = \dfrac{(\pi D_f)^{(1-D_\tau)/2}[4(2-D_f)]^{(1+D_\tau)/2}}{32(3+D_\tau-D_f)} \left(\dfrac{\phi}{1-\phi}\right)^{(1+D_\tau)/2} R_{max}^2$ | Xu and Yu (2008) |
| CPA model | $k_s = \dfrac{d_c^2}{cF}$ | Ghanbarian, (2020a) |

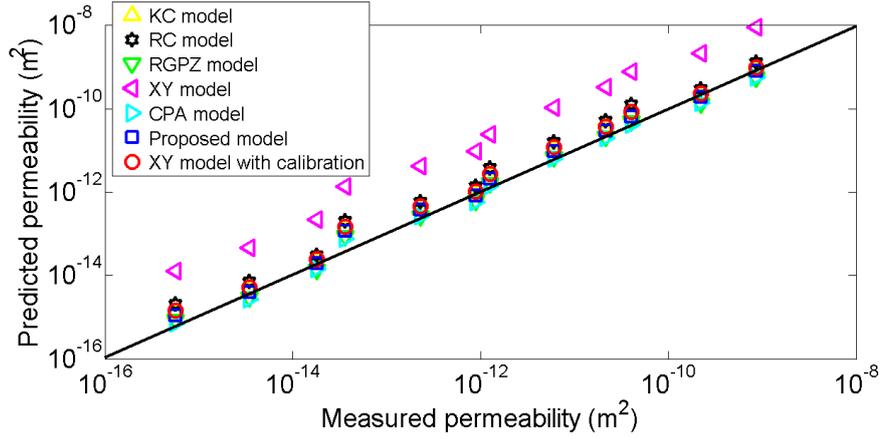

Figure 4: Comparison between measured permeability by Glover and Dery (2010), the proposed model and other models available in literature. The sample properties are given in Table 1. The solid line represents the 1:1 line.

Figure 4 shows the comparison between the predictions of various models from literature and the proposed model given by Eq. (22) for a data set of glass beads reported by Glover and Dery (2010). Sample properties are also shown in Table 1 for the glass beads of Glover and Dery (2010). Table 2 lists some of the models available for the grain-size-based permeability estimation: the RC model proposed by Revil and Cathles (1999), the KC model proposed by Kozeny (1927), the RGPZ model proposed by Glover et al. (2006), the XY model proposed by Xu and Yu (2008) based on the fractal theory and the CPA model proposed by Ghanbarian (2020a) using the critical path analysis. The common values for $m$ and $a$ in the RC model and the RGPZ model are taken to be 1.5 and 8/3 for glass beads, respectively (e.g., Sen et al., 1981; Glover et al., 2006). In the CPA model, $d_c$ is the critical pore diameter that is related to the grain diameter $d$ by $d_c = 0.42d$ and $c$ is a constant coefficient that is equal to 72.2 (e.g., Ghanbarian, 2020a). Due to the similarity between the samples of Glover and Dery (2010) and those reported in Fig. 3 (they are all made up of glass beads or sands), we use $\alpha = 0.009$ as an average optimized value in Table 1. Values of $D_f$, $D_\tau$ are predicted from Eq. (5) and Eq. (6) in the same manner as previously mentioned (see superscript p in Table 1). The root-mean-square deviation (RMSD) for the proposed model, KC model, RC model, RGPZ model, XY model and CPA model is calculated to be $18 \times 10^{-12}$ m², $13 \times 10^{-12}$ m², $118 \times 10^{-12}$ m², $89 \times 10^{-12}$ m², $2351 \times 10^{-12}$ m² and $85 \times 10^{-12}$ m², respectively. The representative comparison shows that the proposed model provides a remarkably good prediction with experimental data reported by Glover and Dery (2010) and with those predicted from the other models. Note that the XY model gave a worse result than others with $R_{max}$ is predicted from Eq. (30). However, the prediction from the XY model could be much improved by calibrating $R_{max}$ a factor 1/3 (dividing $R_{max}$ by a factor 3) as shown by circle symbols in Fig.4 (RMSD = 40×10$^{-12}$ m$^2$). It suggests that the proposed model and the XY model that are related to $R_{max}$ could be improved by calibrating $R_{max}$ predicted from Eq. (30) by a certain factor.

Eq. (22) is also tested in Fig. 5 for a large data set of permeability measurements on similar grain size sediments of different porosity from Chilindar (1964) using the same approach as performed for Fig. 4. The average grain diameter $d$ = 235 μm is taken from Revil and Cathles (1999) for the fine-grained sandstone. Additionally, the KC model, RC model and RGPZ model are also used to explain experimental data reported by Chilindar (1964) and to compare with the proposed model. For the RC model and RGPZ model, $m$ is taken as 1.7 as proposed by Revil and Cathles (1999). For the proposed model, we determine the formation factor from porosity using $F = \phi^{-m}$ with $m$ = 1.7. Feng et al., (2004) and Wei et al., 2015 analyzed the best fit regression parameter $\alpha$ to find $D_f$ using Eq. (5) for natural and artificial porous media from different studies. They found that $\alpha$ = 0.001 gives the best estimate of $D_f$. Additionally, we obtained the maximum pressure ($P_{max}$ = 22.6 MPa) and the minimum pressure ($P_{min}$ = 0.009 MPa) from the capillary pressure measurement of Li and Horne (2006a) for a Berea core sample. Applying the Young–Laplace equation as performed in Ghanbarian et al. (2017b), we obtained $\alpha = R_{min}/R_{max} = P_{max}/P_{min} \approx 0.0004$ that is approximately the same order as 0.001. Therefore, $\alpha$ = 0.001 is rather relevant for consolidated samples and applied in this work.

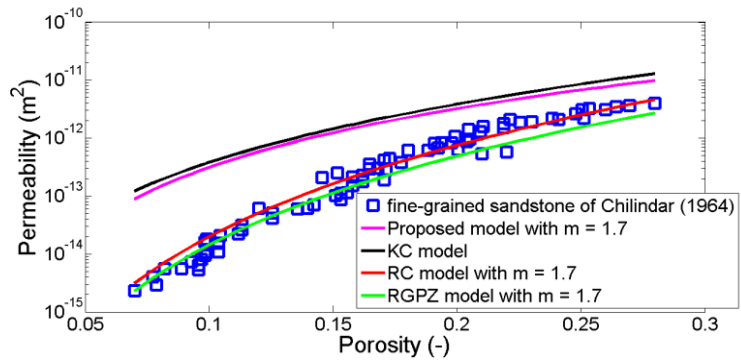

Figure 5: Variation of permeability with porosity for the fine-grained sandstones obtained from Chilindar (1964) (see symbols). The proposed model given by Eq. (22) with $\alpha$ = 0.001, $S_{wirr}$ = 0 and $F = \phi^{-2.2}$ and other ones are used for the prediction.

The calculated RMSD for the proposed model, KC model, RC model and RGPZ model is 68.4×10$^{-14}$ m$^2$, 104×10$^{-14}$ m$^2$, 7.2×10$^{-14}$ m$^2$ and 14.8×10$^{-14}$ m$^2$, respectively. It is seen that the proposed model can reproduce the main trend of experimental data but less accurate than the RC model and the RGPZ model. The reason may be that Eq.

(30) for determining $R_{max}$ from $d$ works quite well for unconsolidated samples that are made up of mono-sized spherical grains as shown in Fig. 3 or Fig. 4. However, for the consolidated samples of sandstone, the rock texture consists of mineral grains of various shapes and sizes and its pore structure is extremely complex. Therefore, Eq. (30) may not be suitable. In this case, one can estimate $R_{max}$ by measuring the capillary pressure and then using the Young–Laplace equation (e.g., Ghanbarian et al., 2017b) or using the micro-CT images and nuclear magnetic resonance measurements (e.g., Daigle, 2016). Another reason may be due to the variation of $\alpha$ from sample to sample (here we use, $\alpha = 0.001$ for all samples).

**3.2 Relative permeability**

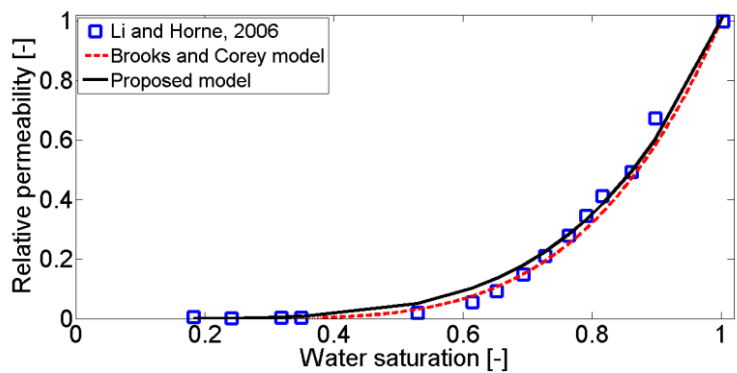

Figure 6: Variation of the relative permeability with water saturation. The symbols are experimental data from Li and Horne (2006b) for Berea sandstone. The solid and dashed lines are predicted from Eq. (21) with $D_f = 1.4$, $D_\tau = 1.05$ and $\alpha = 0.001$ and the model of Brooks and Corey (1964) with $\lambda = 1.9$, respectively.

Figure 6 shows the variation of the relative permeability for wetting phase with water saturation experimentally obtained from Li and Horne (2006b) for a plug of Berea sandstone (see symbols). Eq. (21) is applied to predict the variation of $k_r^w$ with $S_w$ (see solid line). The irreducible water saturation $S_{wirr}$ is reported to be 0.18 (Li and Horne, 2006b). Using the same approach as applied for Fig. 3, we obtain $D_f = 1.4$ and $D_\tau = 1.05$. Note that $\alpha$ is taken as 0.001 for all consolidated rocks in this work as previously mentioned. Additionally, the model of Brooks and Corey (1964) $k_r^w = \left( \dfrac{S_w - S_{wirr}}{1 - S_{wirr}} \right)^{3+2/\lambda}$ with $\lambda = 1.9$ (best fit) is also used to explain experimental data (see dashed line). The calculated RMSD for the proposed model and the model of Brooks and Corey are 0.0043 and 0.0041, respectively. The proposed model is in a very good agreement with experimental data and prediction from the model of Brooks and Corey (1964).

Figure 7 shows the variation of the $k_r^w$ with $S_w$ from different sources. The symbols are experimental data and the solid lines are predicted from Eq. (21). Fig. 7 (a) is obtained from data in Cerepi et al. (2017) for the Brauvilliers limestone with model parameters: $D_f$ = 1.1, $D_\tau$ = 1.05, $\alpha$ = 0.001, $S_{wirr}$ = 0.28 and for the LS2 dolostone with model parameters: $D_f$ = 1.3, $D_\tau$ = 1.05, $\alpha$ = 0.001, $S_{wirr}$ = 0.37. Fig. 7 (b) is obtained from data in Mahiya (1999) for the fired Berea core sample with model parameters: $D_f$ = 1.3, $D_\tau$ = 1.05, $\alpha$ = 0.001, $S_{wirr}$ = 0.29. Fig. 7 (c) is obtained from data in Jougnot et al. (2010) for two Callovo-Oxfordian clay-rock samples with model parameters: $D_f$ = 1.3, $D_\tau$ = 1.05, $\alpha$ = 0.001, $S_{wirr}$ = 0.23. It should be noted that all values for irreducible water saturation $S_{wirr}$ mentioned above are taken from corresponding sources (Cerepi et al., 2017; Mahiya, 1999; Jougnot et al., 2010). It is seen that the model can provide a rather good prediction of the variation of the relative permeability with water saturation.

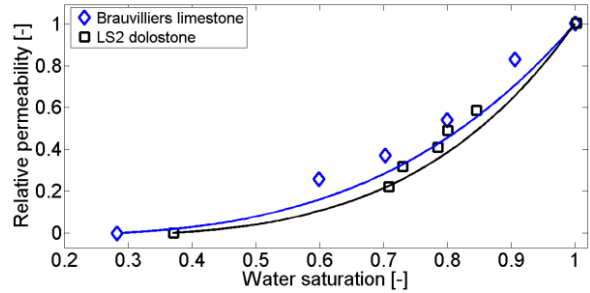

(a)

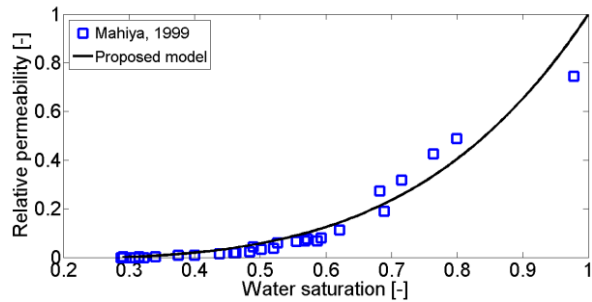

(b)

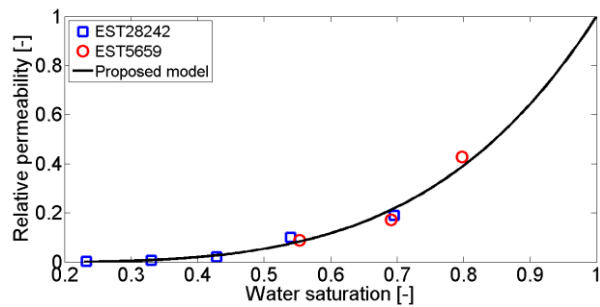

(c)

Figure 7: Variation of the relative permeability with water saturation. The solid lines and symbols are predicted lines and experimental data from: (a) Cerepi et al. (2017) for Brauvilliers limestone ($D_f = 1.1$, $D_\tau = 1.05$, $\alpha = 0.001$, $S_{wirr} = 0.28$) and LS2 dolostone ($D_f = 1.3$, $D_\tau = 1.05$, $\alpha = 0.001$, $S_{wirr} = 0.37$); (b) Mahiya (1999) for a fired Berea core sample ($D_f = 1.3$, $D_\tau = 1.05$, $\alpha = 0.001$, $S_{wirr} = 0.29$); (c) Jougnot et al. (2010) for two Callovo-Oxfordian clay-rock samples ($D_f = 1.3$, $D_\tau = 1.05$ for both, $\alpha = 0.001$, $S_{wirr} = 0.23$).

## 4. Conclusions

We propose a new model to predict the permeability of porous media saturated by one or two fluids based on a bundle of capillary tubes model and the fractal theory for porous media. The model is related to microstructural properties of porous media (fractal dimension for pore space, fractal dimension for tortuosity, porosity, maximum radius, ratio of minimum pore radius and maximum pore radius), water saturation and irreducible water saturation. By comparison with 111 samples of uniform glass bead and sand packs in literature, we show that the proposed model estimated the saturated permeability very well from sample properties. The proposed model is also compared to existing and widely used models from the literature. These results show that the proposed model is in good agreement with the others. The main advantage of the proposed model is that the input parameters are physically-based parameters. Therefore, it may provide an insight into the dependence of the saturated permeability ($k_s$) and the relative permeability ($k_r^w$) on the microstructural parameters of the porous media and it may reveal more mechanisms affecting the $k_s$ and $k_r^w$ than other models. Additionally, the model prediction for the relative permeability has been successfully validated using experimental data for the consolidated media in literature.

**Acknowledgments**

This research is funded by Thuyloi University Foundation for Science and Technology under grant number TLU.STF.19-08.